\documentclass[10pt]{article}%
\usepackage[utf8]{inputenc}%
\usepackage[english]{babel}%
\usepackage[dvips]{graphicx}%
\usepackage{hhline}%
\usepackage{longtable}%
\usepackage{amsmath}%
\usepackage{amssymb}%
\usepackage{caption2}%
\usepackage{multicol}%
\renewcommand{\thefootnote}{\ifcase\value{footnote}\or(*)\or
(**)\or(***)\or(****)\fi}
\begin{document}

\twocolumn[\protect{%
\begin{center}
{\bf \large NEARBY GROUPS OF GALAXIES IN THE HERCULES--BOOTES CONSTELLATIONS}\\
\bigskip
{\large{}I.\,D.\,Karachentsev$^{1,\,}$\footnotemark,
O.\,G.\,Kashibadze$^{1}$,
V.\,E.\,Karachentseva$^{2}$\\}
\bigskip
{\footnotesize\itshape$^{1}$Special Astrophysical Observatory of Russian AS, Nizhnij Arkhyz 369167, Russia\\
$^{2}$Main Astronomical Observatory, National Academy of Sciences of Ukraine, Kiev, 03143 Ukraine\\}
\bigskip
(Received March 2, 2017; Revised March 17, 2017)
\bigskip
\end{center}

\begin{quote}
We consider a sample of 412 galaxies with radial velocities $V_{\rm
LG} < 2500$~km\,s$^{-1}$ situated in the sky region of
${\rm RA}=13^h\hspace{-0.4em}.\,0$--$19^h\hspace{-0.4em}.\,0$, ${\rm
Dec}=+10^{\circ}$...$+40^{\circ}$ between the
Local Void and the Supergalactic plane. One hundred and eighty-one of
them have individual distance estimates. Peculiar velocities of the
galaxies as a function of Supergalactic latitude SGB show signs of
Virgocentric infall at  $SGB < 10^{\circ}$ and motion from the Local
Void at $SGB > 60^{\circ}$. A half of the Hercules--Bootes galaxies
belong to 17 groups and 29 pairs, with the richest group around
NGC\,5353. A typical group is characterized by the velocity
dispersion of $67$~km\,s$^{-1}$, the harmonic radius of $182$~kpc,
the stellar mass of $4.3 \times10^{10} M_{\odot}$ and the
virial-to-stellar mass ratio of $32$.  The binary galaxies have the
mean radial velocity difference of $37$~km\,s$^{-1}$, the projected
separation of $96$~kpc, the mean integral stellar mass of $2.6\times
10^9 M_{\odot}$ and the mean virial-to-stellar mass ratio of about
$8$. The total dark-matter-to-stellar mass ratio in the considered
sky region amounts to $37$ being almost the same as that in the Local
Volume.
\end{quote}

{\bf{}Keywords}: galaxies: kinematics and dynamics---galaxies: distances and
redshifts---galaxies: groups}

\vspace{1cm}]

\section{Introduction}

Mass radial-velocity measurements of galaxies in the recent optical
and radio sky surveys like the SDSS~\cite{aba2009:Karachentsev_n},
HIPASS~\cite{kor2004:Karachentsev_n,wong2006:Karachentsev_n,sta2016:Karachentsev_n},
and ALFALFA~\cite{gio2005:Karachentsev_n,hay2011:Karachentsev_n} led
to a significant enrichment in our understanding of the large-scale
structure and galaxy motions in the nearby universe. Based on the
data on approximately $10^4$~galaxies with radial velocities relative
to the centroid of the Local Group $V_{\rm LG}< 3500$~km\,s$^{-1}$,
Karachentsev and
Makarov~\cite{kar2008:Karachentsev_n,mak2009:Karachentsev_n,mak2011:Karachentsev_n}
compiled catalogs of galaxy systems of different multiplicity
throughout the sky, with the total number of about a thousand. For
galaxy clustering, a new algorithm was applied that took into account
individual masses (luminosities) of galaxies. Using the mutual
separations, radial velocities and luminosities of galaxies in the
$K$-band, they determined virial and stellar masses of galaxy systems
in the  volume of a $48$~Mpc radius, which covers the entire Local
Supercluster and its nearest neighborhood.


One of  important results of these studies was the estimation of the
bulk matter density, enclosed in the systems of galaxies,
$\Omega_m^{\rm vir}=0.08\pm0.02$, which proved to be 3 to 4 times
lower than the average global matter density
$\Omega_m=0.26\pm0.02$~\cite{bac2014:Karachentsev_n}. Various
assumptions were made to explain this discrepancy,  listed
in~\cite{kar2012:Karachentsev_n}:
\begin{list}{}{
\setlength\leftmargin{2mm} \setlength\topsep{2mm}
\setlength\parsep{0mm} \setlength\itemsep{2mm} } \item a) groups and
clusters are surrounded by extended dark haloes, and their main dark
mass is localized outside the virial radius of the system; \item b)
the considered volume of the Local Universe  is not representative,
being located in a giant cosmic void; \item c) a large part of   dark
matter in the universe is not enclosed in groups and clusters, but is
distributed between them in diffuse large-scale structures
(filaments, knots).
\end{list}

\setcaptionmargin{5mm}
\captionstyle{normal}
   \begin{figure*}[]
\centerline{\includegraphics[width=\textwidth]{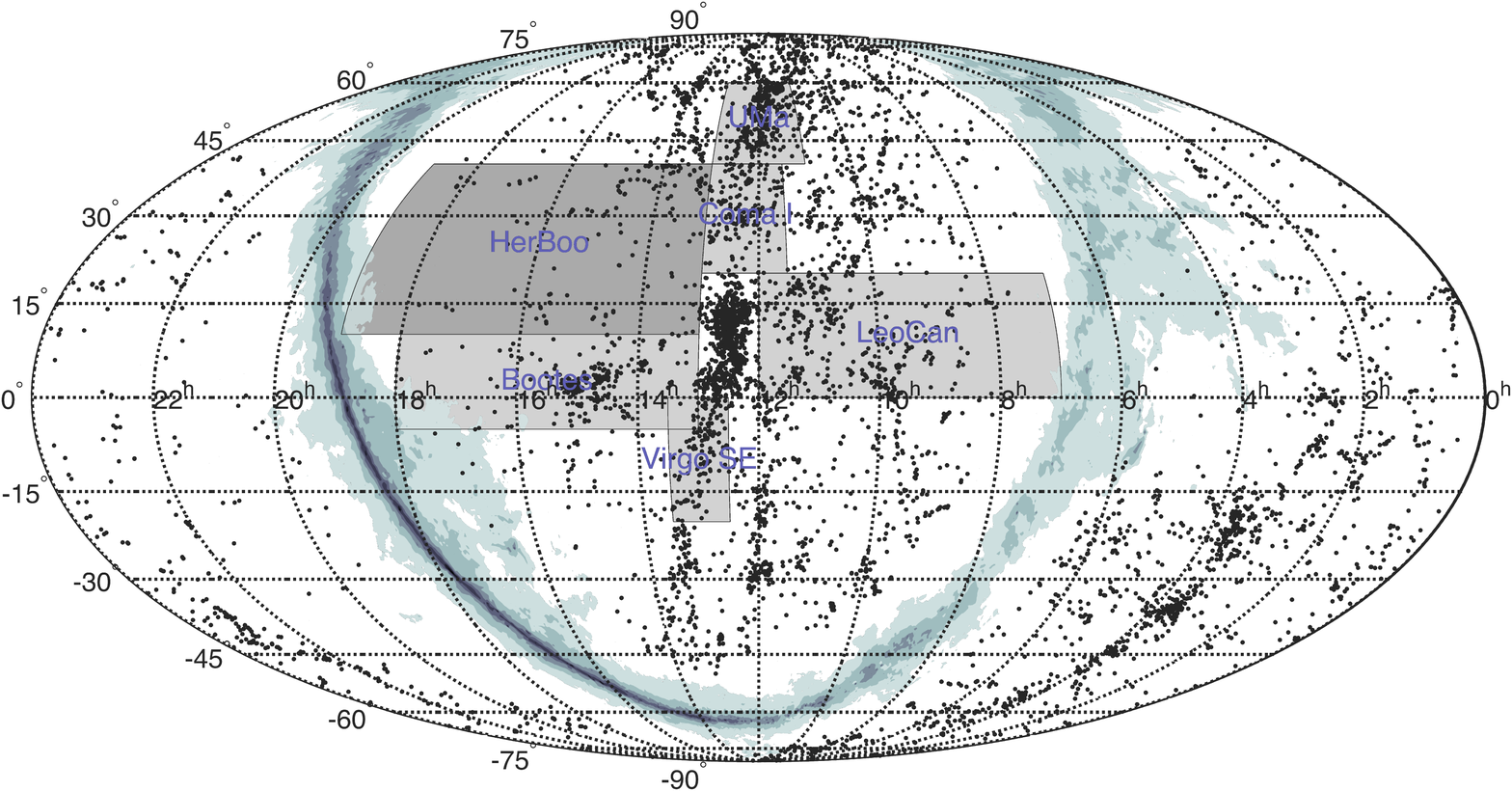}}
\caption{Sky distribution of the Local Supercluster galaxies in
equatorial coordinates. The Hercules--Bootes  and five other regions
we have previously studied   are marked in dark color. A ring-shaped
patchy band indicates a zone of strong extinction.}
   \end{figure*}

However, none of these assumptions has yet received any convincing
observational evidence. Further accumulation of data on the radial
velocities and distances of galaxies is obviously required, as well
as the verification of the effectiveness of the  criterium unifying
the galaxies into groups. Following this idea, we conducted an
analysis of observational data on galaxies in 5~areas with a fairly
representative number of objects. Three of them are located along the
plane of the Local Supercluster to the
north~\cite{kar2011:Karachentsev_n,kar2013a:Karachentsev_n} and to
the south~\cite{kar2013b:Karachentsev_n} from the Virgo cluster. Two
other: the Bootes region~\cite{kar2014a:Karachentsev_n} and the
Leo--Cancer region~\cite{kar2015:Karachentsev_n}
 cover the zones at high supergalactic latitudes.
Location of the studied sky regions is demonstrated in Fig.~1 in
equatorial coordinates. Points mark the positions of 5725 galaxies
with radial velocities of \mbox{$V_{\rm LG}<3000$}~km\,s$^{-1}$, a
ragged ring-shaped band denotes a zone of strong extinction in the
Milky Way. In addition to the 5 previous zones (marked in a light
gray tone), we are considering here a new region, bounded by the
following coordinates: RA from $13^h\hspace{-0.4em}.\,0$ to
$19^h\hspace{-0.4em}.\,0$, and Dec from $+10^{\circ}$ to $+40^{\circ}$.
Since the relative number of galaxies with
distance estimates  steeply drops with increasing radial velocity of
galaxies, we confined ourselves to considering only the objects with
the velocities of $V_{\rm LG}<2500$~km\,s$^{-1}$, which corresponds
to a somewhat larger volume than those in our  previous studies.

\section{Observational data}

The region in question partially overlaps with the SDSS, HIPASS and
ALFALFA sky survey regions. The main source of galaxy data we used is
the NASA Extragalactic Database
(NED)\footnote{\tt{}http://ned.ipac.caltech.edu} with additions from
the HyperLEDA~\cite{mak2014:Karachentsev_n}. Each object with a
radial velocity estimate $V_h$ was visually inspected, and a large
number of false `galaxies' with radial velocities of around zero was
discarded. For many galaxies, we have refined the morphological types
and integral $B$\mbox{-}magnitudes. In the absence of photometric
data the apparent magnitudes of a number of galaxies, usually dwarf
ones, were estimated
 comparing them with the images of other objects having a similar
structure and reliable photometry.

\begin{table*}
\setcaptionmargin{0mm}\onelinecaptionstrue \captionstyle{normal}
 \caption{Galaxies in the Hercules--Bootes region}
 \medskip
 \begin{tabular}{l|c|r|l|r|r|c|l|l}\hline
\multicolumn{1}{c|}{Name}        & RA  (2000.0)  Dec & $V_{\rm LG}$ &
\multicolumn{1}{c|}{T} & \multicolumn{1}{c|}{$B_T$} & $W_{50}$   &
$(m-M)$ & Method & Group
\\
\hline \multicolumn{1}{c|}{(1)}& (2)& \multicolumn{1}{c|}{(3)} &
\multicolumn{1}{c|}{(4)}& \multicolumn{1}{c|}{(5)}&
\multicolumn{1}{c|}{(6)} &\multicolumn{1}{c|}{(7)}& \multicolumn{1}{c|}{(8)} & \multicolumn{1}{c}{(9)} \\
\hline
 UGC\,8085           & 125817.3+143325 & 1971 & Scd  & 14.5   & 204  & 32.37 & tf  & N\,4866\\
 NGC\,4866           & 125927.1+141016 & 1911 & Sb   & 12.14  & 514  & 32.21 & tf  & N\,4866\\
 NGC\,4880           & 130010.6+122900 & 1293 & S0a  & 13.12  &      &       &     &\\
 AGC\,233925         & 130015.7+292859 & 2482 & dIr  & 18.08  & 90   &       &     &\\
 AGC\,239030         & 130018.9+293305 & 970  & dIr  & 18.7   & 40   & 31.57 & bTF &\\
 AGC\,233574         & 130022.1+125525 & 1838 & dIr  & 17.6   & 70   & 32.74 & TF  & N\,4866\\
 UGC\,08114          & 130025.0+134013 & 1909 & Sm   & 15.85  & 130  & 32.91 & TF  & N\,4866\\
 PGC\,1876816        & 130242.8+294458 & 2520 & dIm  & 18.30  &      &       &     &\\
 AGC\,732482         & 130336.8+243132 & 816  & Sm   & 16.2   & 97   & 32.15 & TF  &\\
 AGC\,233930 *       & 130402.7+281833 & 674  & dIm  & 17.17  & 98   &       &     &\\
 KK\,181             & 130433.8+264627 & 1916 & dIr  & 16.88  & 86   & 32.74 & TF  &\\
 KUG\,1302+329       & 130439.4+324054 & 2372 & BCD  & 16.6   &      &       &     &\\
 SDSS\,J130440       & 130440.0+184439 & 766  & BCD  & 17.73  &      &       &     &\\
 IC\,4171            & 130518.8+360610 & 1026 & Sdm  & 15.9   & 89   & 31.73 & TF  &\\
 UGC\,08181          & 130524.6+325400 & 900  & Sdm  & 15.59  & 84   & 31.30 & TF  &\\
 IC\,4178            & 130541.5+360103 & 1215 & dIm  & 16.53  & 64   & 31.48 & TF  &\\
 NGC\,4961             & 130547.6+274400 & 2525 & Scd  & 13.7   & 216  & 33.00 & tf  & N\,4961\\
 IC\,4182            & 130549.6+373618 & 357  & Sm   & 12.00  & 35   & 28.36 & cep & N\,4736\\
 BTS\,165            & 130549.8+274240 & 2516 & dIr  & 17.0   &      &       &     & N4961\\
 AGC\,230077         & 130623.3+102600 & 841  & dIm  & 15.66  & 46   & 30.79 & TF  &\\
 KK\,183             & 130642.5+180008 & 1496 & dIr  & 17.90  & 75   & 31.98 & bTF &\\
 AGC\,230084         & 130656.0+144826 & 915  & dIm  & 16.39  & 49   & 30.40 & TF  &\\
 PGC\,2134801        & 130717.2+384321 & 2423 & dIm  & 17.1   &      &       &     &\\
 AGC\,239031         & 130812.3+290517 & 822  & dIr  & 18.3   & 23   & 29.96 & TF  &\\
 AGC\,742775 *       & 130828.4+200202 & 1430 & dIr  & 18.2   & 146  &       &     &\\
 PGC\,1958740        & 130936.9+314034 & 1449 & BCD  & 17.8   &      &       &     &\\
 AGC\,742788 *       & 131000.8+185530 & 2365 & BCD  & 18.1   & 157  &       &     &\\
 UGC\,08246          & 131004.9+341051 & 825  & SBc  & 14.82  & 116  & 30.90 & tf  & U\,8246\\
 2MFGC\,10495        & 131024.2+213434 & 2547 & Sc   & 16.28  &      &       &     &\\
 $[$MU\,2012$]$\,J13 & 131029.2+341413 & 873  & dIm  & 17.4   &      &       &     & U\,8246\\
 NGC\,5002           & 131038.2+363804 & 1125 & Sm   & 14.69  & 90   & 30.42 & tf  & N\,5005\\
 PGC\,2089756        & 131051.1+365623 & 1061 & dIr  & 17.8   &      &       &     &\\
 NGC\,5005           & 131056.3+370333 & 983  & Sb   & 10.54  & 490  & 31.54 & tf  & N\,5005\\
 UGC\,08261          & 131101.0+353008 & 881  & Sm   & 16.36  & 94   & 31.70 & bTF & N\,5005\\
 SDSS\,J131115       & 131115.8+365912 & 992  & dIr  & 18.43  &      &       &     &\\
 PGC\,2097739        & 131126.8+371843 & 998  & dIm  & 17.48  &      &       &     &\\
\hline
\end{tabular}
\end{table*}

In total, there are 412 galaxies in this region of the sky possessing
radial velocities of  \mbox{$V_{\rm LG}\leq2500$}~km\,s$^{-1}$. Their
list is presented in Table~1, the full version of which is available
in the electronic form in the VizieR
database\footnote{\tt{}http://cdsarc.u-strasbg.fr//viz-bin/qcat?J/other/AstBu/72.2}.
The columns of the table contain: (1)~the name of the galaxy or its
number in well- known catalogs; (2)~equatorial coordinates for the
epoch (2000.0); (3)~radial velocity (in~km\,s$^{-1}$) relative to the
centroid of the Local Group  with the apex parameters accepted in the
NED; (4)~morphological type of the galaxy  according to the
de~Vaucouleurs classification; (5)~integral apparent magnitude of the
galaxy in the $B$-band; (6)~the FWHM of the 21~cm radio line
(in~km\,s$^{-1}$); (7)~the distance modulus corrected for the
extinction in the Galaxy~\cite{sch2011:Karachentsev_n}  and internal
extinction~\cite{ver2001:Karachentsev_n}; (8) the method by which the
distance modulus is determined; (9) the name of the brightest galaxy
in the group/pair to which this galaxy belongs according
to~\cite{mak2009:Karachentsev_n,mak2011:Karachentsev_n}
or~\cite{kar2008:Karachentsev_n}.


The considered region of the sky contains only one galaxy with a
high-accuracy measurement of distance by the cepheids (``cep''). For
four early-type   galaxies the distances are determined by the
surface brightness fluctuations
(``sbf'',~\cite{ton2001:Karachentsev_n}), and for nine very nearby
galaxies the distances are measured by the luminosity of the red
giant branch  (``rgb''). In the remaining galaxies of our sample the
distance moduli are determined by the Tully--Fisher
relationship~\cite{tul1977:Karachentsev_n} with the calibration
according to~\cite{tul2009:Karachentsev_n}:

$$M_B=-7.27(\log W^c_{50}-2.5)-19.99, $$
where $M_B$ is the absolute magnitude in the $B$\mbox{-}band, and the
width of the HI line (in km\,s$^{-1}$)  is corrected for the
inclination of the galaxy.   These estimates are denoted in Table~1
as ``TF''. For late-type galaxies (dIr, dIm, Sm) in which the
HI-value \mbox{$m_{21}=-2.5\log F($HI$)+17.4$} is brighter than the
apparent magnitude, $m_{21}<B$,   we introduced the ``baryon
correction'', replacing the $B$-magnitude in the distance modulus by
$m_{21}$. We designated these cases  as ``bTF''. Galaxies with the
average distance modulus values  from the NED are marked in column
(8) by lower-case letters ``tf''. In total in the region considered,
there are 167 galaxies with distance estimates by  Tully--Fisher,
among them our new estimates make up about 70\%.


In some galaxies from the ALFALFA
HI-survey~{\cite{gio2005:Karachentsev_n,hay2011:Karachentsev_n} the
width of the $W_{50}$ radio line does not correspond to the
structural type of the galaxy  $T$ and its  apparent magnitude $B_T$.
  The reason for this discrepancy may be a confusion
during the  optical identification of a radio source, or a low
signal-to-noise ratio in the HI line. Such galaxies are marked  in Table~1 with an asterisk.

\begin{figure}[]
\setcaptionmargin{5mm}
\captionstyle{normal}
\centerline{\includegraphics[angle=0, width=\columnwidth,clip]{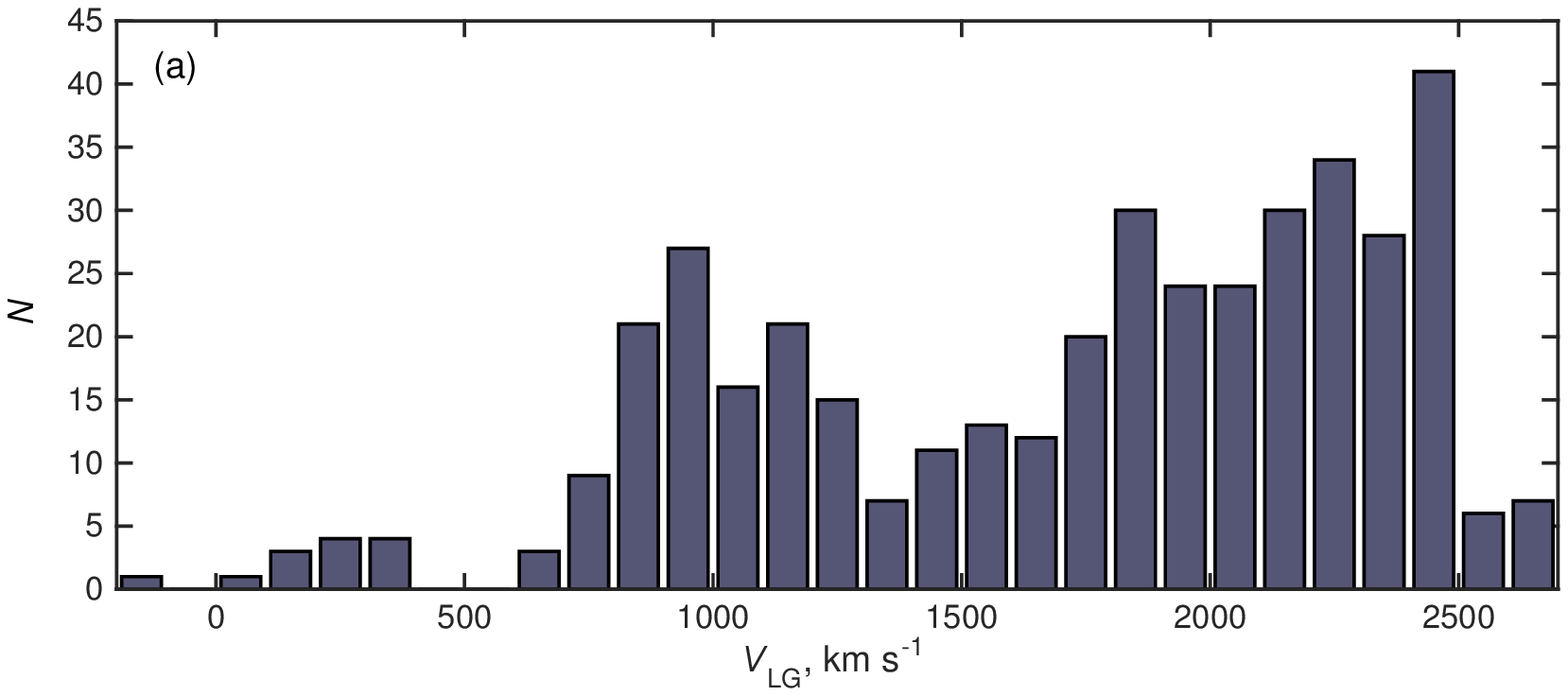}}
\centerline{\includegraphics[angle=0, width=\columnwidth,clip]{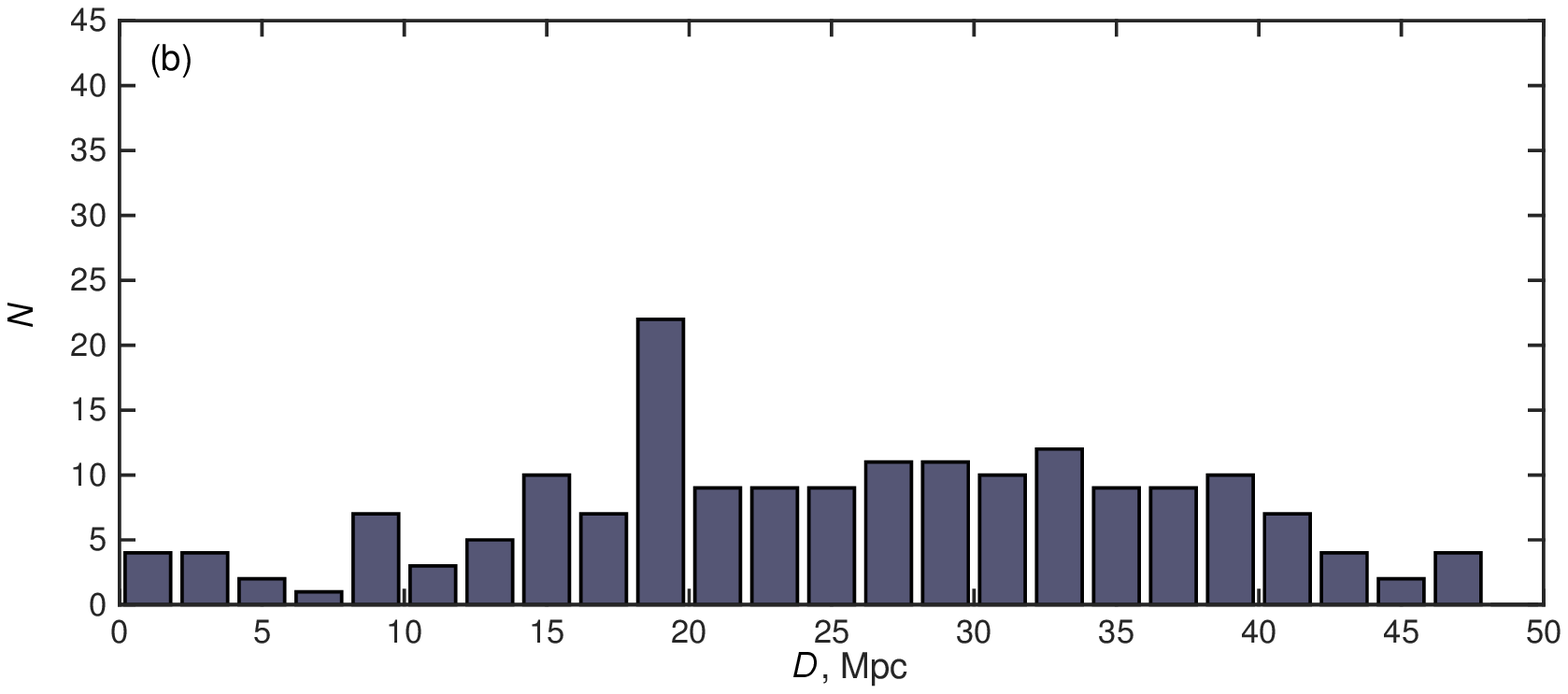}}
\centerline{\includegraphics[angle=0, width=\columnwidth,clip]{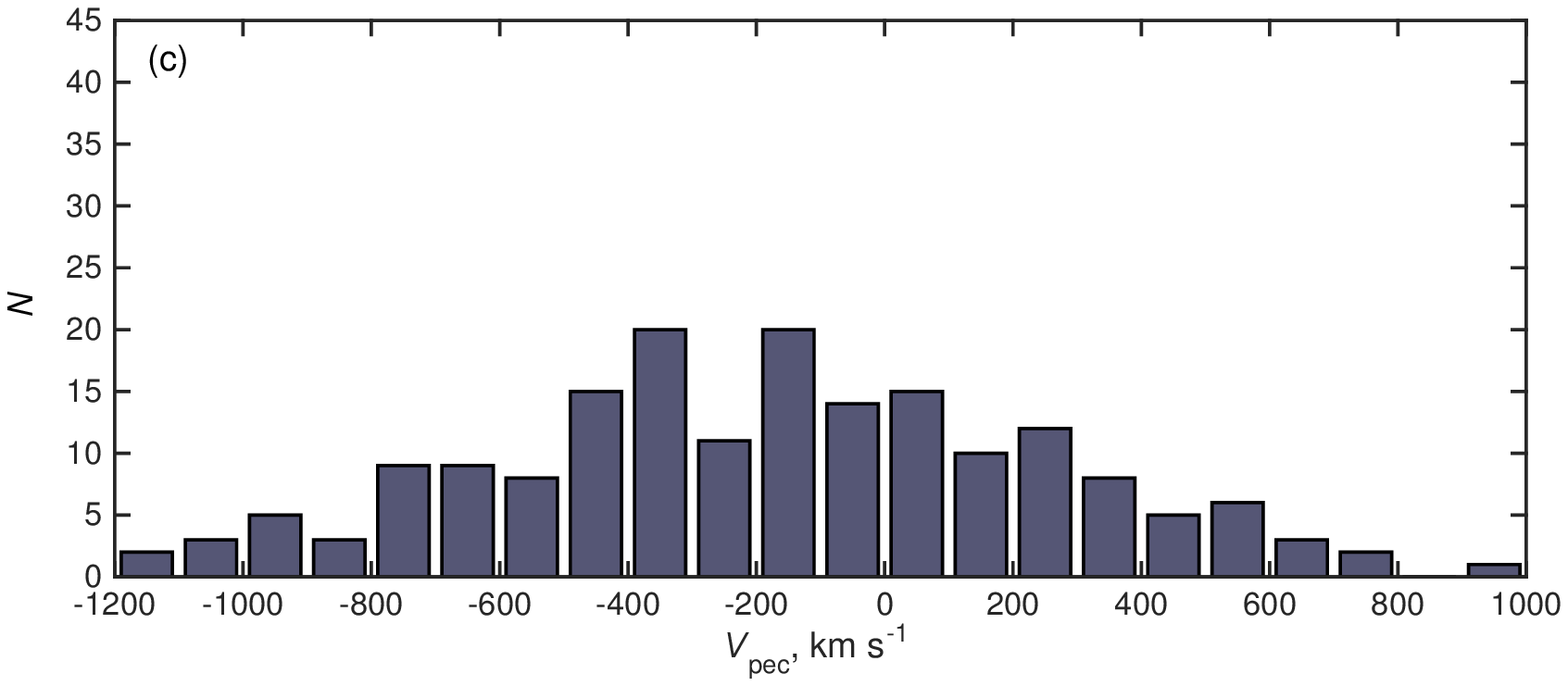}}
   \caption{The distribution of the number of galaxies in the Hercules--Bootes region based on (a)~radial velocities relative to the centroid of the Local Group, (b)~distances and (c)~peculiar velocities.}
\end{figure}

The distribution of 412 galaxies by radial velocity and 181 galaxies
by the distance are shown in Figs.~2a, b, respectively. Several
galaxies with velocities greater than $2500$~km\,s$^{-1}$ belong to
the  NGC\,5353 and NGC\,6181 group members, the average velocity of
which lies at the boundary of the selected range  $V_{\rm LG}$.
Figure~2a reveals a local excess of galaxies with the velocities of
about~$1000$~km\,s$^{-1}$, which is obviously  due to the presence of
galaxies, associated with the spurs of the Virgo cluster in the
considered region. A part of these galaxies apparently  causes a peak
that is noticeable in the $N(D)$ distribution at the distance of
\mbox{$D\simeq18$~Mpc}~(Fig.~2b).


Figure~2c reproduces the distribution of   181 Hercules--Bootes galaxies
 throughout the peculiar velocity   $V_{\rm pec} =V_{\rm
LG}-H_0\times D$,  given the value of the Hubble parameter of
$H_0=73$~km\,s$^{-1}$Mpc$^{-1}$.   The histogram has a completely
symmetrical shape with the average value of $V_{\rm pec} = -179
$~km\,s$^{-1}$ and the variance of $425$~km\,s$^{-1}$.
 With an average distance of the sample galaxies
of about~$26$~Mpc and a typical distance error of around 20\%, the
expected accuracy of peculiar velocity   is $380$~km\,s$^{-1}$. The
excess of the observed velocity dispersion   over the expected
dispersion may indicate the existence of large-scale motions of
galaxies in the  Hercules--Bootes region. Note that in the
  volume we consider, the number of galaxies with    peculiar velocity estimates
is approximately twice as large as that contained in the Cosmicflows-2
and Cosmicflows-3
databases of galaxy distances~\cite{tul2013:Karachentsev_n,tul2016:Karachentsev_n}.

 \begin{figure*}[]
 \setcaptionmargin{5mm}
\captionstyle{normal} \centerline{\includegraphics[angle=0,
width=0.9\textwidth,clip]{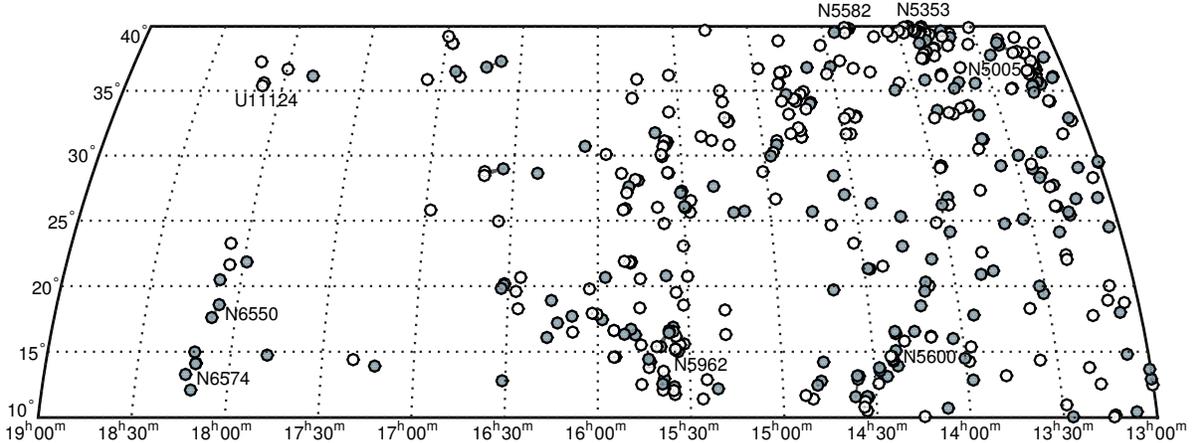}} \caption{The
distribution of galaxies in the  Hercules--Bootes region in
equatorial coordinates. Galaxies with distance estimates are
represented by filled circles, the remaining galaxies with velocities
of $V_{\rm LG}<2500$~km\,s$^{-1}$  are shown by open circles. The
brightest members of some groups are marked with their names.}
\end{figure*}

The total distribution of 412  Hercules--Bootes galaxies in equatorial
coordinates is shown in Fig.~3. Galaxies with distance estimates
and without them are marked by dark and bright circles, respectively.
The most populated groups are designated by the names of their brightest members. This diagram demonstrates the clustering of galaxies into systems of different
multiplicity, as well as a global increase in the number density   of galaxies from
the left  to the right edge approaching the equator of the Local
Supercluster.

\section{Groups and pairs of galaxies}

Combining the galaxies into systems of different multiplicity, we were guided by
the criterion proposed in the paper~\cite{mak2011:Karachentsev_n}. According to it, each  virtual pair $ij$ must satisfy the condition of negative total energy
$$V_{ij}^2R_{ij}/(2GM_{ij})<1,$$ where $G$ is the gravity constant,
and the condition of finding its components inside the ``zero-velocity sphere'', which
isolates this pair relative to the global Hubble expansion
$$\pi H_0^2 R^3_{ij}/(8GM_{ij})<1,$$
where $H_0$ is the Hubble parameter. Here $V_{ij}$ and $R_{ij}$ are
the differences between the radial velocities and the
projection of  mutual separations of the virtual pair components,
$M_{ij}$ is their total mass expressed through the
$K$-luminosity $M/L_K=\kappa M_{\odot}/L_{\odot}$. To estimate the total mass of the galaxy, we took the value of the dimensionless parameter
$\kappa=6$, at which the structure and
virial mass of the well-studied nearby groups is best reproduced. The
clustering algorithm involves a sequential revision of all the galaxies
of the initial selection and the subsequent grouping of all pairs possessing
common members.

\begin{table*}
\setcaptionmargin{0mm}\onelinecaptionstrue \captionstyle{normal}
\caption{Properties of the galaxy groups in the  Hercules--Bootes region}
\medskip
\begin{tabular}{l|c|r|r|r|r|r|c|r|c|c|r|r}\hline
\multicolumn{1}{c|}{Group}   &  J2000.0     &   $N_V$&   $\langle V_{\rm LG}\rangle$ &  $\sigma_V$&   $R_h$&    $\log M^*$ &  $\lg M_p$  &  $N_D$ & $ \langle m-M\rangle$ &  $\sigma(m-M)$ &  \multicolumn{1}{c|}{$D$} & \multicolumn{1}{c}{$V_{\rm pec}$}\\
\hline \multicolumn{1}{c|}{(1)}& \multicolumn{1}{c|}{(2)}& (3) &
\multicolumn{1}{c|}{(4)} & \multicolumn{1}{c|}{(5)}&
\multicolumn{1}{c|}{(6)} &\multicolumn{1}{c|}{(7)} & (8)
&\multicolumn{1}{c|}{(9)} &(10)& (11)& (12) &
\multicolumn{1}{c}{(13)}
\\ \hline
NGC\,4736 &    125053.1+410714 & 13 &   352 &  50 &  338 & 10.64 &  12.33 &  13 & 28.28 &  0.14 & 4.5  &  $24$\\
NGC\,4866 &    125927.1+141016 &  5 &  1909 &  58 &  156 & 11.21 &  12.68 &   4 & 32.56 &  0.28 & 32.5 &  $-464$\\
NGC\,5005 &    131056.2+370333 & 13 &  1010 & 114 &  224 & 11.48 &  12.97 &   9 & 31.24 &  0.41 & 17.7 &  $-282$\\
NGC\,5117 &    132256.4+281859 &  4 &  2414 &  27 &  424 & 9.97  &  11.95 &   2 & 32.80 &  0.09 & 36.3 &  $-236$ \\
NGC\,5353 &    135539.9+402742 & 62 &  2593 & 195 &  455 & 12.07 &  13.69 &  16 & 32.73 &  0.39 & 35.2 &  $16$\\
NGC\,5375 &    135656.0+290952 &  3 &  2311 &  47 &  66  & 10.62 &  11.68 &   1 & 32.94 &  ~--  & 38.7 &  $-514$\\
NGC\,5582 &    142043.1+394137 &  6 &  1685 & 106 &  93  & 10.60 &  12.44 &   2 & 31.82 &  0.54 & 23.1 &  $-1$\\
NGC\,5600 &    142349.5+143819 &  6 &  2295 &  81 &  275 & 10.69 &  12.38 &   3 & 32.05 &  0.91 & 25.7 &  $419$\\
UGC\,9389 &    143533.2+125429 &  4 &  1822 &  45 &  204 & 9.68  &  12.08 &   4 & 32.54 &  0.19 & 32.2 &  $-529$ \\
PGC\,55227&    152929.2+260024 &  3 &  2119 &  14 &  17  & 9.21  &  10.05 &   2 & 32.34 &  0.17 & 29.4 &  $-27$\\
NGC\,5961 &    153516.2+305152 &  5 &  1891 &  63 &  86  & 10.14 &  12.20 &   1 & 32.51 &  ~--  & 31.8 &  $-430$\\
NGC\,5962 &    153631.7+163628 &  8 &  1996 &  97 &  60  & 11.23 &  13.01 &   6 & 32.60 &  0.35 & 33.1 &  $-420$\\
NGC\,5970 &    153830.0+121110 &  4 &  1949 &  92 &  141 & 10.81 &  12.54 &   3 & 32.45 &  0.28 & 30.9 &  $-307$\\
UGC\,10043  &    154841.3+215210 &  5 &  2214 &  67 &  65  & 10.37 &  11.88 &   1 & 33.03 &  ~--  & 40.4 &  $-735$\\
NGC\,6181 &    163221.0+194936 &  4 &  2568 &  53 &  196 & 11.06 &  12.14 &   3 & 32.65 &  0.19 & 33.9 &  $93$\\
UGC\,10445  &    163347.4+285904 &  3 &  1118 &  23 &  230 & 9.92  &  11.60 &   1 & 31.57 &  ~--  & 20.6 &  $-386$\\
NGC\,6574 &    181151.2+145854 &  3 &  2456 &  15 &   70 & 11.08 &  10.71 &   2 & 32.36 &  0.43 & 29.6 &  $295$\\
\hline
\multicolumn{2}{c|}{Average}&  9 &  1924 &  67 &  182&  10.63&  12.14 &   4 & 32.15 & 0.34&  29.2 &  $-$205\\ \hline
\end{tabular}
\end{table*}

Therefore, 17 galaxy groups with populations of three or more members
were selected in the considered region of the sky. The main data
about them are presented in the columns of Table~2: (1) the name of
the brightest member of the group; (2) the equatorial coordinates of
the center of the group; (3) the number of members with measured
radial velocities; (4) the average  radial velocity of the group
(km\,s$^{-1}$); (5)   the root-mean-square velocity of galaxies
relative to the average (km\,s$^{-1}$); (6) the harmonic average
radius of the group (kpc); (7)   the logarithm of the total stellar
mass of the group (in    $M_{\odot}$ units), determined from the
$K$-band luminosity of its members   at
\mbox{$M^*/L_K=M_{\odot}/L_{\odot}$}; (8)
 the logarithm of the projected
(virial) mass in  $M_{\odot}$ units:
  $$M_p=(32/\pi G)(N-3/2)^{-1} \sum^N_{i=1} \Delta V^2_i R_i,$$
where $\Delta V_i$ and $R_i$
  are   the radial velocity and the projection
distance of the $i$-th galaxy relative to the center of the system;
(9) the number of members with measured distances; (10)   the average
distance modulus of the group; (11)   the variance of the moduli of
the group members; (12)    linear distance in Mpc at the mean modulus
$\langle m-M\rangle$; (13)  peculiar velocity of the center of the
group, $V_{\rm pec} = \langle V_{\rm LG}\rangle-73 \langle D\rangle$,
(km\,s$^{-1}$). The last row of the table corresponds to the average
values  of the parameters.


As follows from these data, the characteristic radius of the group
($182$~kpc) and the characteristic dispersion of  radial velocities
($67$~km\,s$^{-1}$) prove to be typical of the Local Group and other
nearby groups in the Local Volume~\cite{kar2014b:Karachentsev_n}. The
characteristic stellar mass of the group in Table~2, $M^*\simeq
4\times 10^{10} M_{\odot}$, and virial-to-stellar mass ratio,
\mbox{$M_p/M^*\simeq 32$}, are also typical for the well-studied
nearby groups.

\begin{table*}
\setcaptionmargin{0mm}\onelinecaptionstrue \captionstyle{normal}
\caption{Pairs of galaxies in the  Hercules--Bootes region}
\medskip
\begin{tabular}{l|r|r|c|r|r|r|c} \hline
 \multicolumn{1}{c|}{Name}    & $\langle V_{\rm LG}\rangle$ &  $\Delta V$  &   $D$   &  $R_p$\, &  $\log M^*$&  $\log M_{\rm orb}$&    $\Delta(m-M)$\\
\hline \multicolumn{1}{c|}{(1)}& \multicolumn{1}{c|}{(2)}&
\multicolumn{1}{c|}{(3)} & \multicolumn{1}{c|}{(4)}& \multicolumn{1}{c|}{(5)} & \multicolumn{1}{c|}{(6)} & \multicolumn{1}{c|}{(7)}& \multicolumn{1}{c}{(8)} \\
\hline
 UGC\,8246    & 849   & 48   &  15.1 & 27  &  8.48  &  10.86 & --\\
 UGC\,8318    & 2417  &  18  &  41.5 &  56 &  9.71  &  10.22 & --\\
 AGC\,732599  & 1902  &  24  &  26.1 &  55 &  8.13  &  10.57 & --\\
 UGC\,8507    & 980   &  29  &  12.6 & 132 &  9.25  &  11.11 & 0.87\\
 PGC\,169748  & 728   &  13  &  14.6 &  77 &  7.87  &  10.18 & --\\
 UGC\,8667    & 1417  &  26  &  19.4 &  13 &  8.63  &  10.02 & --\\
 NGC\,5303    & 1473  &  22  &  18.7 &  15 &  9.81  &  9.93  & --\\
 IC\,4341   & 2386  &  39  &  38.4 & 103 &  10.06 &  11.26 & --\\
 NGC\,5611    & 2076  &  80  &  25.2 &  54 &  10.23 &  11.61 & --\\
 UGC\,9274    & 1162  &  12  &  15.4 &  42 &  8.82  &  9.85  & 0.10\\
 IC\,1014   & 1278  &   2  &  18.3 &  73 &  9.37  &  8.53  & 0.43\\
 UGC\,9320    & 864   &  8   &  12.6 & 133 &  7.08  &  9.99  & --\\
 UGC\,9356    & 2181  &  56  &  35.0 &  61 &  9.76  &  11.35 & 0.96\\
 NGC\,5727    & 1578  &   4  &  23.5 &  55 &  9.20  &  9.01  & --\\
 UGC\,9504    & 1592  &  11  &  21.8 &   8 &  8.98  &  9.05  & --\\
 UGC\,9519    & 1711  &  18  &  23.4 & 101 &  10.00 &  10.58 & --\\
 NGC\,5762    & 1798  &  7   &  29.1 & 218 &  9.92  &  10.10 & 0.32\\
 PGC\,2080256 & 1978  &   1  &  27.1 &  10 &  8.70  &  7.07  & --\\
 UGC\,9562    & 1334  & 112  &  18.3 &  21 &  9.19  &  11.49 & --\\
 NGC\,5798    & 1881  &  24  &  25.1 & 152 &  9.98  &  11.01 & --\\
 AGC\,733735  & 2100  &   42 &  38.3 & 106 &  8.90  &  11.34 & --\\
 NGC\,5958    & 2119  &   12 &  29.0 &  43 &  10.17 &  9.86  & --\\
 NGC\,5956    & 1905  &   70 &  26.1 & 143 &  10.59 &  11.91 & --\\
 NGC\,6012    & 2012  &  175 &  21.7 &  48 &  10.50 &  12.24 & --\\
 UGC\,10086   & 2378  &  166 &  32.6 &   9 &  10.18 &  11.44 & --\\
 NGC\,6207    & 1035  &    4 &  17.5 & 360 &  10.16 &  9.83  & 0.28\\
 NGC\,6255    & 1100  &   23 &  19.9 & 194 &  9.60  &  11.08 & --\\
 UGC\,10625   & 2256  &    1 &  33.3 & 16  &  9.03  &  7.27  & --\\
 NGC\,6550    & 2410  &   15 &  24.2 & 454 &  10.85 &  11.08 & 0.21\\
\hline
 Average  & 1686  &   37 &  24.3 &  96 &   9.42 &  10.34  &    0.45\\ \hline
\end{tabular}
\end{table*}

\begin{figure*}[]
\setcaptionmargin{5mm}
\captionstyle{normal}
\centerline{\includegraphics[angle=0, width=0.8\textwidth,clip]{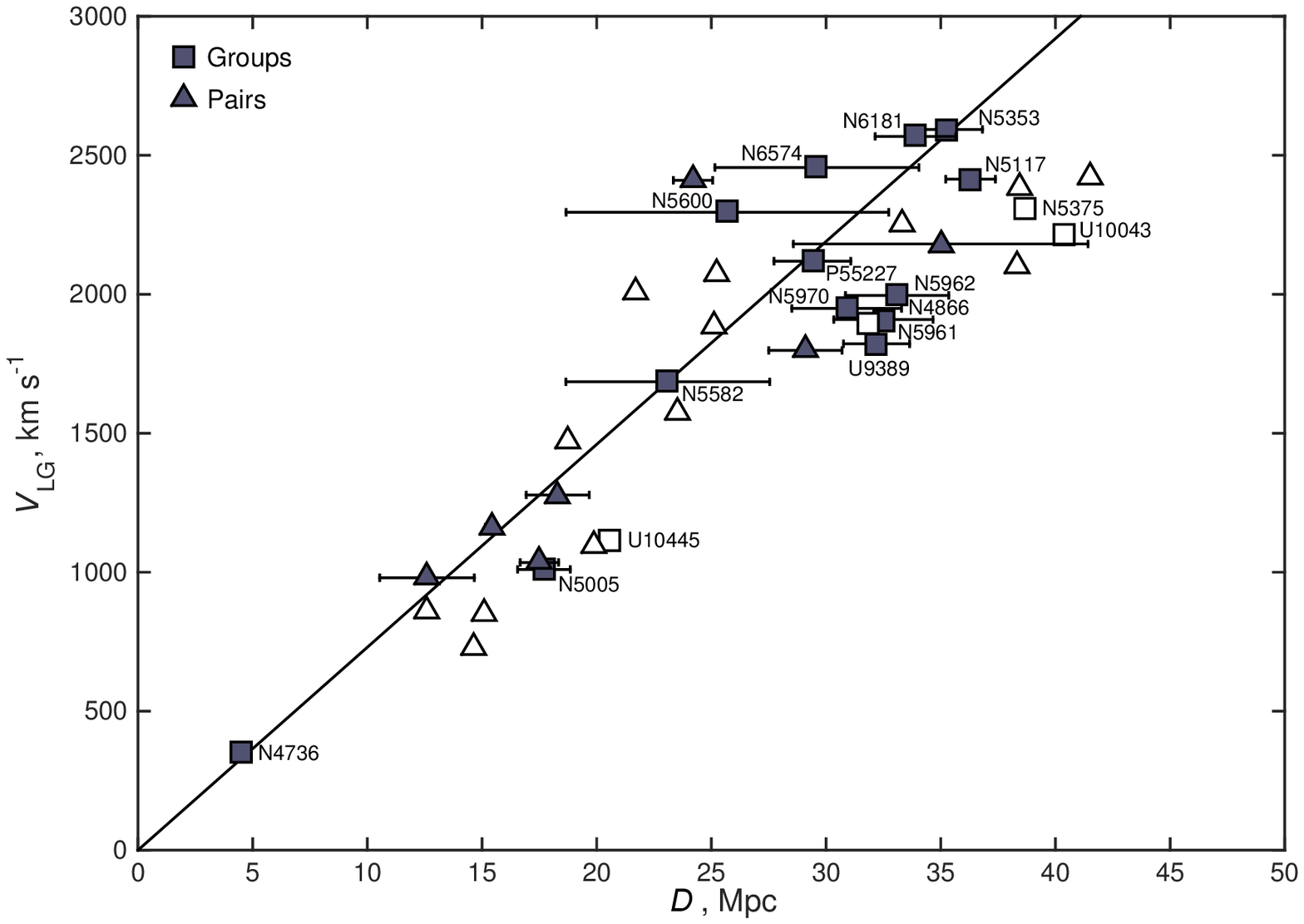}}
 \caption{Hubble  velocity--distance diagram for group centers
(squares) and pairs of galaxies (triangles). Systems with distance estimates for two
and more members are indicated by solid symbols with mean errors marked.}
\end{figure*}

If the clustering algorithm for galaxies is  correctly selected, then
the distance modulus variance of group members should be determined
by the distance measurement errors. In our case, the distances of
most galaxies are measured by the Tully--Fisher method, the error of
which is approximately 20\% or 0.4. The average modulus variance
for members of 17 groups amounts to  0.34, i.e. it is in
agreement with the expected value.


Among the 17 groups listed in Table~2, a group of galaxies around
NGC\,5353 is distinguished by large stellar   and virial
masses. The structure and morphological composition of this group were
  investigated in~\cite{tul2008a:Karachentsev_n}.   From the radial velocities of fifteen brightest members of the group the authors have identified the virial mass of this system as $2.1\times 10^{13}M_{\odot}$.
Our estimate of the total mass of the  NGC\,5353 group from 62 galaxies with measured velocities gives a twice higher value.
In this case, the  $M_p/M^*=47$ ratio for it also
looks typical of rich groups, similar to the nearby group Leo\,I.
Considering the filamentary structures of galaxies in the broad vicinity of the
Virgo cluster, Kim  et al.~\cite{kim2016:Karachentsev_n} suggested   that the
NGC\,5353 group is connected to the Virgo by a long (about $25^{\circ}$) thin
filament. However, our data on the velocities and distances of galaxies in
this area does not support this assumption.


In addition to 17 groups, this region contains 29 pairs of galaxies,
a summary on which is presented in Table~3. The designations of the
columns in it are similar to the previous table. A typical pair has a
difference of the component radial velocities of \mbox{$\langle
\Delta V_{12}\rangle=37$~km\,s$^{-1}$}, the projection distance
between the components  of $\langle R_{12}\rangle = 96$ kpc and the
stellar mass of $\langle \log M^*/M_{\odot}\rangle=9.42$. The
projected (orbital) mass of the pair,$$M_p=(16/\pi G) (\Delta
V_{12})^2 R_{12}$$ is on the average  8 times the total stellar mass:
$\langle \log (M_p/M^*) \rangle=0.92$. The mean difference of the
distance moduli for the components of the pairs,  0.45,
testifies to an insignificant share of fictitious optical pairs among
them.


Figure~4 shows the Hubble   velocity--distance  diagram for the
centers of groups and pairs of galaxies in the Hercules--Bootes
region. Groups with individual distance estimates for two or more
members are denoted by solid squares, while the groups with   $N_D=1$
are shown as empty squares. The pairs of galaxies with    $N_D=2$ and
$N_D=1$ are depicted, respectively, by solid and empty triangles. The
straight line corresponds to the Hubble parameter of
$73$~km\,s$^{-1}$Mpc$^{-1}$. It is clear from these data   that an
increase in the number of group members    with Tully--Fisher
distance estimates favours to reduce dispersion of peculiar
velocities in the group centers. We expect that groups of galaxies
identified by the criterion~\cite{mak2011:Karachentsev_n} with
\mbox{$N_D>4$}  have a typical error of the average distance
measurement of about~10\%, i.e. their average velocities and average
distance estimates     by the
  Tully--Fisher method can be successfully used for
tracing   the field of peculiar velocities along with other
high-accuracy methods (``cep'', ``SN'', ``rgb'').

\section{Peculiar motions   in the   Hercules--Bootes region}
 \begin{figure*}[]
 \setcaptionmargin{5mm}
\captionstyle{normal} \centerline{\includegraphics[angle=0,
width=0.9\textwidth,clip]{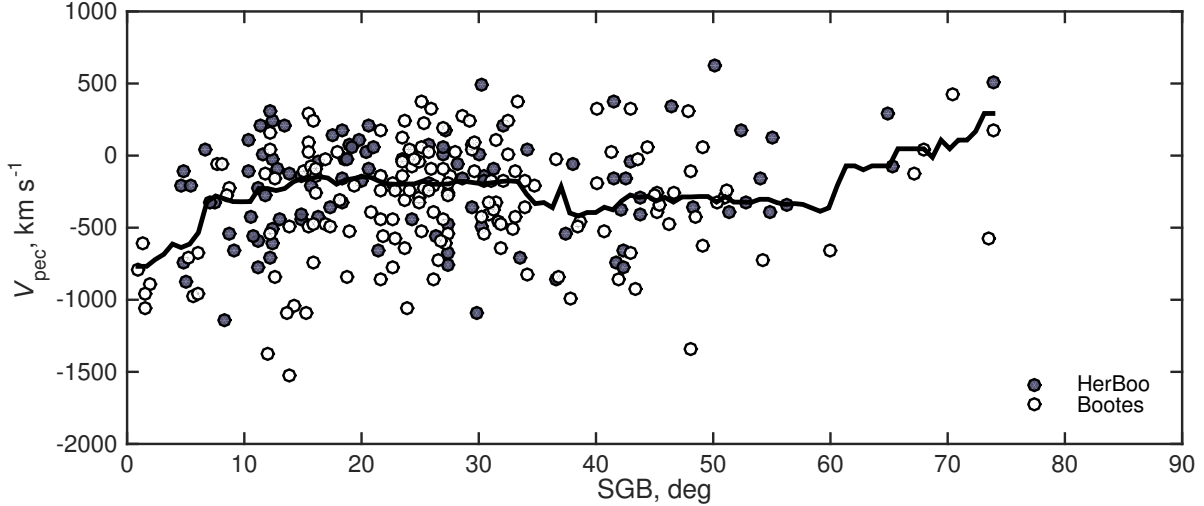}} \caption{Peculiar
 velocity of the  Hercules--Bootes  (filled circles) and Bootes strip galaxies
(open circles) depending on the supergalactic latitude. The broken
line corresponds to a running median with a 2.5 window.}
 \end{figure*}

 Considering the observational data on radial velocities and distances
of the galaxies in the Local Supercluster and its environs, Tully et
al.~\cite{tul2008b:Karachentsev_n} have identified two main factors,
forming the local field of peculiar velocities: the infall of
galaxies to the center of the Virgo cluster (the so-called
Virgocentric infall) at a characteristic rate of
about~$180$~km\,s$^{-1}$ and the outflow of galaxies from the center
of the expanding Local Void with the typical velocity of
about~$260$~km\,s$^{-1}$. It is obvious that both these effects
should influence the  peculiar velocity field  in the
Hercules--Bootes region, stretching between the Local Void and the
Virgo cluster.


Since the suspected center of the Local  Void  is located near the
northern supergalactic pole at the    latitude SGB$\simeq
+77^{\circ}$~\cite{tik2006:Karachentsev_n}, and the center of Virgo is almost on the equator of the
Local Supercluster  (SGB~$\simeq-2^{\circ}$), then the approximate
  orthogonality of these directions facilitates the analysis of the   peculiar
velocity field between them.


For all the galaxies in the \mbox{RA $=$ [$13^h\hspace{-0.4em}.\,0,
19^h\hspace{-0.4em}.\,0$]}, \mbox{Dec $=$ [$+10^{\circ}, +40^{\circ}$]} 
region with the  $D$ and
$V_{\rm pec}$ estimates we have determined supergalactic coordinates
SGL and SGB. To make the picture complete we added 161 galaxy from
the Bootes strip to this sample: \mbox{RA $=$ [$13^h\hspace{-0.4em}.\,0,
18^h\hspace{-0.4em}.\,0$]},
Dec $=$ [$-5^{\circ}, +10^{\circ}$], with radial velocities of $V_{\rm
LG}<2000$~km\,s$^{-1}$. The behavior of the median value of  peculiar
velocity of galaxies along the supergalactic latitude is shown in
Fig.~5. The solid broken line corresponds to a running median with an
averaging window of 2.5.
 In order to
 make the combined sample more homogeneous, we excluded from it the
  Hercules--Bootes galaxies with   \mbox{$V_{\rm LG}>2000$}~km\,s$^{-1}$.
The galaxies of the Hercules\mbox{--}Bootes and Bootes~strip regions
are depicted in the figure by solid and open circles, respectively.


As follows from these data, at intermediate  supergalactic latitudes of
 \mbox{SGB $=$
[$+10^{\circ}, +60^{\circ}$]}  the values of the median peculiar velocity of galaxies
vary in a narrow range \mbox{from~$-200$} to \mbox{$-400$~km\,s$^{-1}$}.
 At low
supergalactic latitudes, \mbox{SGB $< 10^{\circ}$}, the median value
$V_{\rm pec}$ drops  to the minimum value of about  \mbox{$-700$~km\,s$^{-1}$.}
Most of the galaxies in the \mbox{SGB $< 10^{\circ}$}  zone have the distances of
$D > 16$~Mpc, i.e. they are behind the Virgo cluster.
Falling in the direction of the
Virgo cluster as a massive local attractor, these galaxies acquire
a significant negative line of sight peculiar velocity.
The observed amplitude of the flow  to Virgo
proves to be comparable with the virial dispersion of the cluster
velocities  \mbox{$\sigma_V\simeq 650$~km\,s$^{-1}$}.


On the other side of the diagram at  \mbox{SGB $> 60^{\circ}$} the statistics
of the  peculiar velocity data is poor. Nevertheless, there is a  tendency of growth
 of the galaxy median velocity   to the region of positive values.
 The interpretation of this effect depends on the model assumptions on the
structure and kinematics of the Local Void.
If its center is located at the
distance of  \mbox{$D_c \simeq 10$~Mpc}~\cite{nas2011:Karachentsev_n} at
\mbox{SGB$_c = +77^{\circ}$},   then the galaxies with a typical distance of
\mbox{$D \simeq 26$~Mpc}  around the expanding void will have a
positive component of the  line of sight  peculiar velocity.
However,
the real configuration of the Local Void according to the data of~\cite{ely2013:Karachentsev_n} looks more complicated. According to these authors, the Local Void  is a chain of empty volumes, which, meandering like a
horseshoe, covers both the Local Volume and the Virgo cluster.


Rizzi et al.~\cite{riz2017:Karachentsev_n} have recently measured
with high accuracy the rgb-distances  to two close dwarf
galaxies located not far from the direction to the center of the Local Void:\\
KK246 ($D=6.95$~Mpc, $SGB= +40^{\circ}$),\\ as well as ALFAZOA1952+1428
($D=8.39$~Mpc, \mbox{$SGB= +76^{\circ}$}).\\  The galaxies have an
average peculiar velocity of $-90\pm24$~km\,s$^{-1}$.  Given the
recession velocity of the Milky Way itself from the center of the
Local Void of about~$230$~km\,s$^{-1}$   this corresponds to the
recession velocity   of these galaxies from the center of the void of
about~$320$~km\,s$^{-1}$. Therefore, four different sets of
observational data
(\cite{tul2008b:Karachentsev_n,nas2011:Karachentsev_n,riz2017:Karachentsev_n},
this article) on   peculiar velocities of galaxies in the vicinity of
the nearest void show that the walls of the void move outward from
its center at a characteristic velocity of several
hundred~km\,s$^{-1}$.

\section{Discussion}

\begin{table*}
\setcaptionmargin{0mm}\onelinecaptionstrue \captionstyle{normal}
\caption{Comparative properties of the three studied  sky regions}
\medskip
\begin{tabular}{l|c|c|c}\hline
\multicolumn{1}{c|}{Parameter}              &   Leo--Cancer &       Bootes strip &       Hercules--Bootes \\
\hline
 Sky area, sq.deg.       &    1477      &     1121     &       2447\\
 $V_{\rm LG}^{\rm max}$,  km\,s$^{-1}$         &    2000     &      2000     &       2500\\
 Volume,  Mpc$^3$          &    3084      &     2337     &       9975\\
 $N_V$                     &     543      &      361    &         412\\
 $N_D$                      &    290      &      161     &       181\\
 Number density, Mpc$^{-3}$  &    0.176     &     0.154    &      0.042\\
 $N$(groups+pairs)           &   23+20      &    13+11     &     17+29\\
 Fraction of isolated         &    0.51      &     0.44     &      0.50\\
$\sum  M^*_{\rm syst}, 10^{12} M_{\odot}$  &    3.50      &     2.63     &      2.62\\
 $\rho^*_{\rm syst}/\langle\rho^*\rangle$        &    2.47      &     2.45     &      0.57\\
 $\sum  M_p, 10^{13} M_{\odot}$     &    9.10      &     8.80      &     9.58\\
 $\sum M_p/\sum M^*$          &     26       &      33      &       37\\ \hline
\end{tabular}
\end{table*}
As it was repeatedly noted
(see~\cite{mak2011:Karachentsev_n,kar2012:Karachentsev_n}), the total
virial mass of groups and clusters in the Local Universe with a
diameter of about $100$~Mpc is only 8\mbox{--}10\% of the critical
density, which is about 3 times smaller than the global density of
dark matter, \mbox{$\Omega_m=0.26\pm0.02$}. A significant enlargement
of the observational base thanks to the recent optical and HI sky
surveys left this contradiction  almost unchanged. In this regard, it
is useful to consider how the problem of missing dark matter looks
like based on the data for various regions of the Local Supercluster.


Table~4 presents the main characteristics of the three regions of the
sky we have studied: Leo\mbox{--}Cancer, the Bootes strip and
Hercules--Bootes, located outside the plane of the Local
Supercluster, which is laden with the   projection effects. The first
three lines of the table contain: the area of  each region in square
degrees, the maximum velocity, up to which the galaxies were
considered, and the volume of each region in Mpc$^3$ at
$H_0=73$~km\,s$^{-1}$Mpc$^{-1}$. The following two lines (4 and 5)
contain the number of galaxies in these zones with   measured radial
velocities $(N_V)$ and distances $(N_D)$. As shown in line 6, the
number densities of galaxies with measured velocities are
approximately the same in the Leo\mbox{--}Cancer and Bootes regions,
while in the Hercules--Bootes zone this density is significantly
lower than in the others. The number of groups and pairs of galaxies
(line 7) varies significantly from zone to zone, while the least
dense area, Hercules--Bootes contains an increased number of pair
systems consisting of low-luminosity galaxies. The relative number of
single (non-clustered) galaxies (line 8) makes up about a half in
each region, along with that,   the field population is dominated by
dwarf galaxies. Line~9 shows the total stellar mass of galaxy members
in systems of different multiplicity. Line~10 contains the value of
stellar density, expressed in relation to the average cosmic density
$\langle\rho^*\rangle= \langle j_K\rangle=4.3\times10^8
M_{\odot}$\,Mpc$^{-3}$ according to~\cite{jon2006:Karachentsev_n} at
$M^*/L_K= M_{\odot}/L_{\odot}$~\cite{bell2003:Karachentsev_n}. As we
can see, the Leo--Cancer and Bootes regions have the  average stellar
density 2.5 times larger than the global density, and the
Hercules--Bootes region is located lower than the level of average
cosmic density. The last two lines present: the total virial
(projected) mass of all groups and pairs, as well as its relation to
the stellar mass sum  of these systems. Note that the    $\sum
M_p/\sum M^*$ ratio varies in a small range from 26 to 37, despite
significant differences in the average stellar density from one area
to another.


Given the global average  stellar matter density of
$\langle\rho^*\rangle= 4.3\times10^8 M_{\odot}$ Mpc$^{-3}$, which is
currently known   with the error of about 30\%, a dimensionless ratio
of critical matter density to stellar density amounts to
$\rho_c/\langle \rho^*\rangle
=350\pm100$~\cite{fuk2004:Karachentsev_n}. The observed values of
$\sum M_p/\sum M^*$ in the three  considered regions prove to be an
order of magnitude smaller than the critical ratio.


Note that we did not take into account single non-clustered galaxies,
which account for about a half of the total number of galaxies in
each region. However, additional analysis shows that their
contribution in the total stellar mass does not exceed 20\%, since
the majority of field galaxies have a low luminosity. In addition,
single galaxies with their dark haloes contribute to both the
denominator and the numerator of the   $M_{\rm DM}/M^*$ ratio. This
is why accounting for single galaxies can not significantly affect
the values  presented in the last row of   Table~4. It is important
to emphasize that the value of    $M_{\rm DM}/M^*\simeq30$  is
typical of the dark haloes of the Milky Way, M\,31, M\,81 and other
brightest galaxies of the Local
Volume~\cite{kar2014b:Karachentsev_n}. Therefore, the observed lack
of virial mass in the nearby systems of galaxies is still an actual
problem for the cosmology of the Local Universe.

{\bf Acknowledgments}

\noindent{}In this paper we used the NASA Extragalactic Database (NED) and
HyperLEDA database, as well as the HIPASS, ALFALFA and SDSS sky
survey data. IDK and OGK are grateful to  the Russian Science
Foundation  for the support  (grant no.~14--02--00965).



\begin{flushright}
{\it Translated by A.~Zyazeva}
\end{flushright}



\end{document}